\newcommand{\ignore}[1]{}
\documentclass[amsmath,amssymb,floatfix,lengthcheck,preprintnumbers,prl,showpacs,superscriptaddress,nofootinbib,twocolumn,aps]{revtex4-1}
\usepackage{float}
\usepackage[left]{lineno}
\usepackage{amsmath}
\usepackage{amssymb}
 \usepackage{graphicx}
\usepackage{braket}
\usepackage{enumitem}
\usepackage{bm}
\usepackage{xcolor} 

\newcommand{\<}{\langle\,}

\renewcommand{\>}{\rangle}

\newcommand\mS{{\mathbb S}}
\newcommand\mR{{\mathbb R}}

\newcommand\mIcos{{\cal I}}
\newcommand\cO{{\cal O}}

\newcommand\be{\begin{equation}}
\newcommand\ee{\end{equation}}
\newcommand\bea{\begin{eqnarray}}
\newcommand\eea{\end{eqnarray}}
\newcommand{\bdm}{\begin{displaymath}}
\newcommand{\edm}{\end{displaymath}}
\newcommand{\bal}{\begin{align}}
\newcommand{\eal}{\end{align}}
\newcommand\nn{\nonumber \\}

\renewcommand\ell{l}

\renewcommand{\eqref}[1]{Eq.~\ref{#1}}

\usepackage{color}
\newcommand\ricci{  \textbf{Ric}}
\usepackage[section]{placeins}
\renewcommand{\footnote}[1]{ }

\begin{document}

\title{Radial Lattice Quantization of 3D $\phi^4$ Field Theory}

\author{Richard C. Brower}
\affiliation{Boston University, Boston, MA 02215}

\author{George T. Fleming}

\affiliation{Yale University, Sloane Laboratory, New Haven, CT 06520}
\author{Andrew D. Gasbarro}

\affiliation{AEC Institute for Theoretical Physics, University\"{a}t Bern, 3012 Bern, Switzerland}
\author{Dean Howarth}

\affiliation{Nuclear Science Division, Lawrence Livermore National Laboratory, Livermore, CA 94550}

\author{Timothy G. Raben}
\affiliation{Michigan State University, East Lansing, MI 48824}

\author{Chung-I Tan}
\affiliation{Brown University, Providence, RI 02912}

\author{Evan S. Weinberg}
\affiliation{NVIDIA Corporation, Santa Clara, CA 95050}

\date{\today}

\begin{abstract}
The quantum extension of classical finite elements, referred to as quantum finite elements
({\bf QFE})~\cite{Brower:2018szu,Brower:2016vsl},  is applied to the  radial quantization of  3d $\phi^4$ theory on a
simplicial lattice for the  $\mathbb R \times \mathbb S^2$ manifold.  Explicit counter terms to cancel the one- and two-loop
ultraviolet defects are implemented to reach the quantum continuum theory.  Using the Brower-Tamayo~\cite{Brower:1989mt} cluster Monte Carlo algorithm,  numerical results support the QFE ansatz  that the  critical conformal field theory (CFT)  is reached in the continuum  with the full isometries of $\mathbb R \times \mathbb S^2$ restored. The Ricci curvature term, while technically irrelevant in the quantum theory, is shown to dramatically
improve the convergence opening, the way for high precision Monte Carlo simulation to determine the CFT 
data: operator dimensions, trilinear OPE couplings and the central charge. 

\end{abstract}
\maketitle

\setlength{\parskip}{.2in}

\paragraph{\bf Introduction --}\label{sec:intro}

There are many quantum field theories on general Riemann manifolds which would benefit from a rigorous extension of lattice field theory methods beyond flat Euclidean space. 
 As a  test, we choose the classic prototype of 3d $\phi^4$ in radial quantization on $\mR \times \mS^{2}$  in comparison with the well studied 3d Ising CFT~\cite{El-Showk:2014dwa}. 
In principle there are many advantages to studying CFTs using radial quantization.
In radial quantization, the CFT is mapped from flat space $\mR^d$ to the cylinder $\mR \times \mS^{d-1}$ with the 
radius  of the sphere fixed to $R$.  Translations, $ t \rightarrow t + t_0$, along the cylinder correspond to   exponential displacements  in the radial distance $r = R\exp[t]$,  allowing one to reach the equivalent of exponential  scales on a finite  lattice.

\begin{figure}[t]
\centerline{\includegraphics[width=0.9\linewidth]{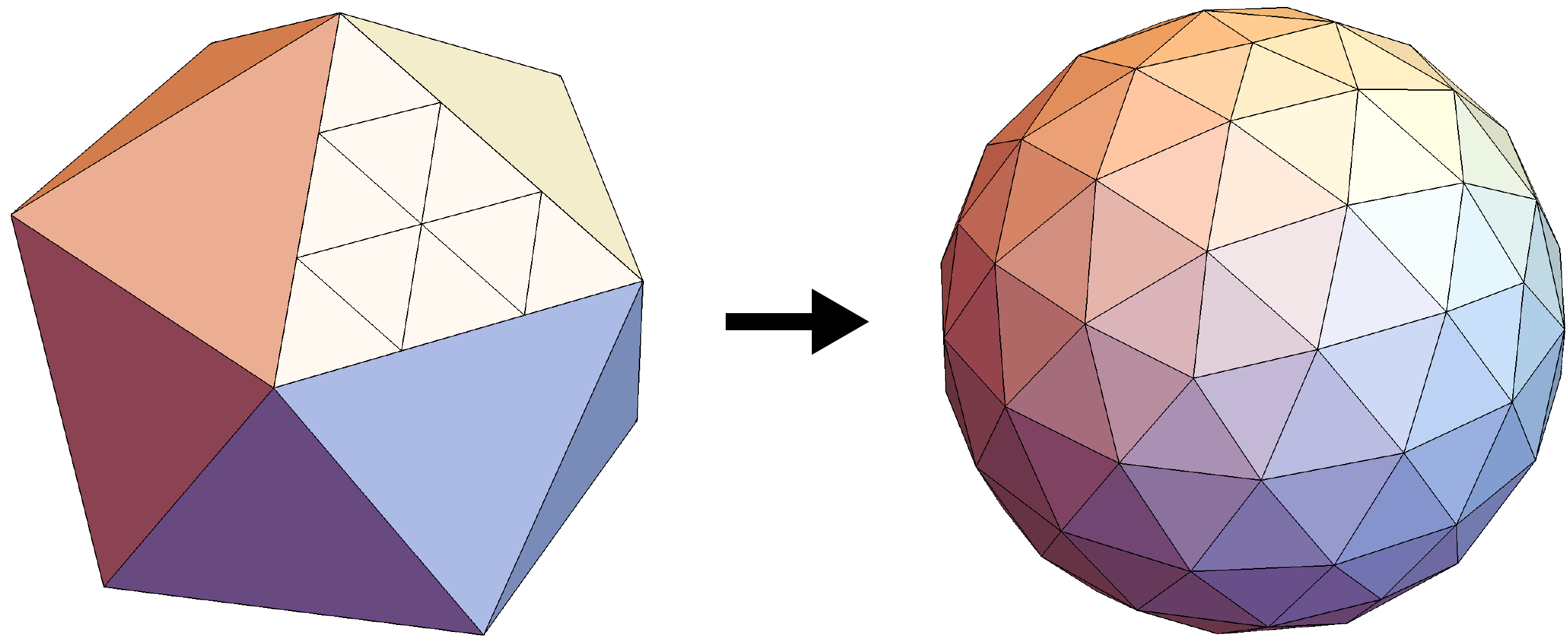}}
\caption{\label{fig:icos} The L-th level refinement of the icosahedron
subdivides triangles  into $L^2$ smaller triangles for a total of $N_\triangle = 20L^2$ faces,  $E = 30L^2$ edges and
$N = 2 + 10L^2$ sites.  Illustrated on the left is the  $L=3$ icosahedral refinement  with $ 2 + 10L^2 = 92$ vertices and on the right subsequently projected onto the unit sphere. }
\end{figure}

However, as pointed out by Cardy~\cite{Cardy:1984rp,Cardy:1985lth},
for $d>2$ a lattice implementation faces severe difficulties,
because there is no uniform sequence of lattices approaching
the spherical manifold, $\mS^{d-1}$.  A first attempt by Brower, Fleming, and Neuberger~\cite{Brower:2012vg} placed the 3d Ising model  on a cylindrical lattice $\mR \times \mIcos$   with the sphere $\mS^2$ approximated 
by a uniformly triangulated icosahedron $\mIcos$ as illustrated on the left side of Fig~\ref{fig:icos}. 
While the results 
were encouraging, not surprisingly a small breaking of spherical
symmetry was observed in the splitting of the $2l +1 $ degenerate rotational multiplets. 
The third descendant ($l = 3$) splits into 2 irreducible multiplets of the icosahedral group, even when extrapolated to the continuum limit. To remove this defect, a new lattice discretization method has been
developed, referred to as quantum finite elements
(QFE)~\cite{Brower:2016vsl}, conjectured to
apply to any renormalizable quantum field theory on a smooth Riemannian manifold. 

The QFE construction begins by defining a series
of refined simplicial lattices, which approach the target manifold, 
and a classical lattice action using the finite element method (FEM)
based on the discrete exterior calculus (DEC). 
While this is sufficient for classical solutions
to the equation of motion, it fails
in the quantum path integral due to sensitivity
to the irregular UV cutoff intrinsic to the simplicial approximation of the target manifold.  To overcome this problem, explicit QFE counter terms are introduced, which we conjecture will restore the exact non-perturbative
quantum physics as the cutoff is removed. 

To date this QFE method has been tested 
with numerical simulations for the 2d $\phi^4$ theory on $\mS^2$ and has been found to be in precise agreement with the exact solution of the minimal $c = 1/2$ Ising CFT~\cite{Brower:2018szu}. The goals of this paper are to test the QFE method   for $\phi^4$ theory on $\mR\times \mS^2$ in comparison with the 3d Ising CFT and to demonstrate its potential to give high precision lattice  Monte Carlo predictions to extend and complement results from the conformal bootstrap~\cite{el2012solving,El-Showk:2014dwa}. 

\begin{figure}[ht]
	\centering
	\includegraphics[width=\columnwidth]{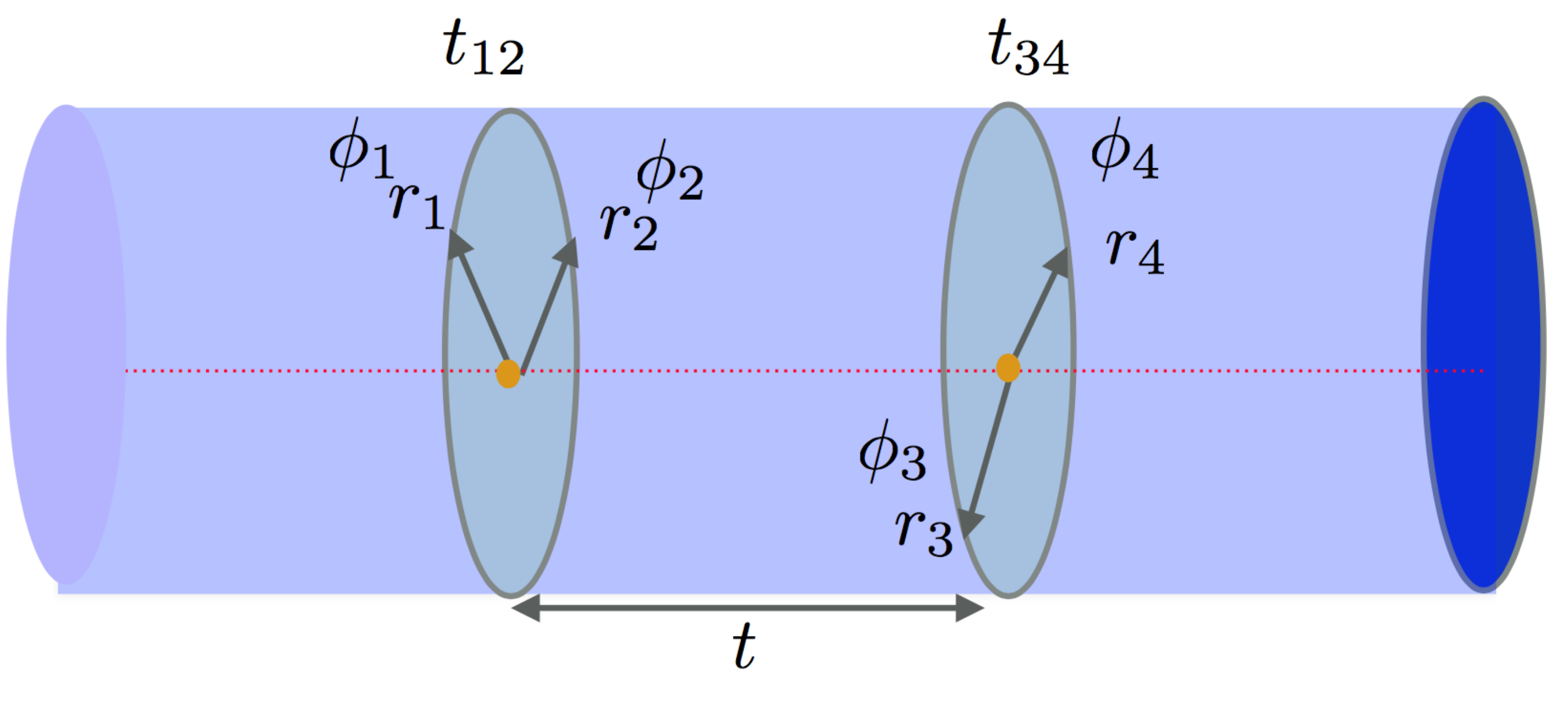}
	\caption{The OPE expansion for the 4 point function on the cylinder is computed by placing the field on two
	spheres separated by $t = |t_{12} - t_{13}|$.\label{fig:4pt}  }
\end{figure}

\paragraph{\bf Classical Simplicial Lattice Action --} \label{sec:qfe}
Conformal field theories have an enlarged symmetry group, promoting the
d-dimensional Euclidean Poincar\'{e} group to the
conformal group $O(d+1,1)$ or the isometries of
the global $AdS^{d+1}$ manifold.   As a consequence,
a CFT can equivalently be  quantized after  a  Weyl transformation 
from a flat Euclidean manifold $\mR^d$ to a cylinder  $\mR \times
\mS^{d-1}$,
\begin{align}
ds^2_{flat} = dx^{\mu}dx^{\mu} &=  r^2[(d\log r)^2 + d\Omega^2_{d-1}]\nn 
\xrightarrow{Weyl} ds^2_{cylinder} &= dt^2 + R^2 d\Omega^2_{d-1} \;
,\label{eq:Radial}
\end{align}
 with a sphere of fixed radius $R$ and flat coordinate
$t =\log(r/R)$ along the length of the cylinder. In radial quantization  the dilatation
operator, which is  conjugate to translations in  {\em radial time, $t$,} plays the role of the Hamiltonian.
The cylinder, illustrated in Fig.~\ref{fig:4pt}, resides on the boundary of $AdS^{d+1}$ space. The problem, as in all lattice
constructions, is to construct a  discrete lattice 
action which  is rigorously equivalent to the continuum
 quantum path integral on the target manifold as the cutoff is removed. Since the radius of the sphere provides 
 an intrinsic IR scale, the  continuum limit for
 the conformal theory is now defined
 by the limit, $a/R \rightarrow 0$, relative to the UV
 cutoff at  lattice spacing $a$.

We begin by implementing a finite element method discretization of the continuum action,
\begin{align}
S_{cont} =  \frac{1}{2}\int_{\mathcal{M}} d^dx \sqrt g & [ 
g^{\mu \nu} \partial_\mu \phi(x) \partial_\nu \phi(x) + \xi_0  \ricci \; \phi^2(x) \nonumber \\
&+  m^2 \phi^2(x)+ \lambda \phi^4(x)] \; ,  \label{eq:continuum}
\end{align}
for $\phi^4$-theory on a curved Riemann manifold.  The  action 
includes a Ricci scalar term with non-zero
coefficient, $\xi_0 = (d - 2)/(4(d-1))$ for $d \ge 3$.  For our radial
quantization on $\mR \times \mS^{d-1}$, the Ricci term, 
$\ricci = (d -1)(d -2)/R^2$, is a constant determined by the radius of the sphere, and as such represents a  shift in the mass. 
For the massless free theory ($m = \lambda = 0$) in 3d, it is a relevant
marginal  operator required by conformality. As
a result the free conformal scalar is gapped with a
spectrum,  $l (l +1) +1/4 = (l + 1/2)^2$,  that is the square of the  spectrum, $l +1/2$, for a free massless Dirac operator. The dimension of the conformal scalar primary and its descendants is $\Delta_{\phi,l}  = 1/2 + l$.

To construct the radial lattice action for $\mR \times \mS^2$, we introduce a sequence of simplicial 
lattice approximations to  $\mS^2$, as illustrated in Fig.~\ref{fig:icos}. The simplicial lattice is refined by introducing finer triangulations on each icosahedral face and projecting onto geodesics on the sphere.  
This spherical lattice is copied  uniformly along the $\mR$ 
cylindrical axis with lattice spacing $a_t$.  It is important to 
carefully introduce the FEM action in physical units relative to
the radius, $R$, of the sphere,  
\begin{align}
&S_{FEM}=  \frac{a_t}{2}  \big[\sum_{y\in\<x,y\>}  \frac{l^*_{xy}}{l_{xy}}  (\phi_{t,x} - \phi_{t,y} )^2
    +   \frac{\sqrt{g_{x}}}{4 R^2} \phi^2_{t,x} \nonumber \\
    &+   \sqrt{g_{x}} [\frac{ (\phi_{t,x} - \phi_{t+1,x} )^2}{a^2_t} +m^2  \phi^2_{t,x}
     +   \lambda   \phi^4_{t,x}] \big] \; , \label{eq:FEMphysicalUnits}
\end{align}
with the Einstein summation convention for  $x = 1,2,\cdots, N$
for sites on each sphere and  $t = 1, 2,\cdots, L_t$ along the length of the cylinder  with periodic boundary conditions. 

This form follows from the elegant discrete exterior calculus (DEC) implementation of the FEM on a simplicial complex
and its Vorono\"i dual. For the special case of a 2d triangulation the DEC form happens to coincide with piecewise linear elements, where
$l_{xy}$ are the lengths of the edges of triangles shown on the right side of Fig.\ref{fig:icos}, $l^*_{xy}$ are the lengths of edges between
circumcenters on the dual lattice, and $\sqrt{g_x}$ are the Vorono\"i dual areas at each site.  
On a general n-dimensional  simplicial complex, the resultant DEC Laplace-Beltrami operator is,
\be
*\; {\bf d} * {\bf d} \; \phi_x = \frac{1}{|\sigma^*_0(x)|} \sum_{y\in\<x,y\>}  \frac{|\sigma^*_1(xy)|}{|\sigma_1(xy)|} (\phi_x	-\phi_y)
\ee
with the replacement: $l_{xy} \rightarrow |\sigma_1(xy)|$,
$l_{xy}^* \rightarrow  |\sigma^*_1(xy)|$ and $\sqrt{g_x}  \rightarrow  |\sigma^*_0(x)|$.  The discrete
Hodge Star $(*)$ transfers differential forms between the  simplicial lattice and  Vororo\"i dual
polytopes, weighted by appropriate volume elements.
The reader is referred our  earlier papers~\cite{Brower:2016vsl,
  Brower:2018szu,Gasbarro:2019kgj} and to the vast FEM literature  for details~\cite{Arnold_2010}. 

It is important to appreciate the theoretical consequences of the discrete exterior calculus.  When properly applied, DEC guarantees exact convergence of the simplicial
action Eq.~\ref{eq:FEMphysicalUnits} to the  classical action
Eq.~\ref{eq:continuum}, and therefore all lattice  solutions converge to solutions of the Euler-Lagrange PDE's as the cutoff is removed.  Moreover the DEC formalism extends
naturally to higher dimensions with  higher spin gauge fields and K\"ahler-Dirac or staggered fermions. With the addition of a FEM spin connection, an extension to Wilson lattice fermions, including domain wall with an extra flat direction,  was formulated in Ref.~\cite{Brower:2016vsl}.

In the next step, we follow the standard methods of numerical simulation by rewriting the lattice action in terms of dimensionless fields and parameters: 
\begin{align}
\label{eq:FEMaction}
S &=  \frac{1}{2}  \big[\sum_{y\in\<x,y\>}  \frac{l^*_{xy}}{l_{xy}}  (\phi_{t,x} - \phi_{t,y} )^2
    +   \frac{a^2}{4 R^2}  \sqrt{g_{x}} \phi^2_{t,x} \\
    &+   \sqrt{g_{x}} [\frac{a^2}{a^2_t}(\phi_{t,x} - \phi_{t+1,x} )^2 +m^2_0  \phi^2_{t,x}
     +   \lambda_0   \phi^4_{t,x}] \big]  \; . \nonumber
\end{align}
 On a hypercubic lattice, with a uniform 
 lattice spacing ($a$), this is  equivalent
 to working in units so that $a =1$.  Here it is a bit subtler.  The geometry of the manifold has introduced an explicit  IR length scale through the radius of the sphere and  two UV cutoffs: the longitudinal lattice spacing, $a_t$, and
 a characteristic edge length $a^2$ on the
 sphere.  For convenience, we have defined $a^2$ relative
 to the average area of  Vorono\"i polytope, $ A^* = \< \sqrt{g_x}\>= a^2\sqrt{3}/2$ at each vertex.  Of course
  there are other possible choices such as average
 area of triangles, $A_\triangle \simeq A^*/2 = a^2\sqrt{3}/4$ or the average squared edge length, $\< l^2 _{xy} \> \simeq 0.752 a^2$ which are equivalent to $\cO(a^4/R^2)$.
 
To unambiguously recover the physical scales, we need to provide
our change to dimensionless variables, $\widetilde g_x$ and $\widetilde \phi_{t,x}$, 
\be
\sqrt{g_x} = A^* \sqrt{\widetilde g_x}    \quad ,\quad   \phi_{t,x}  = \widetilde \phi_{t,x}/Z_0 \; .  
\label{eq:Rescaling}
\ee
This fixes the mean lattice measure $\<\sqrt{\widetilde g_x}\>$  exactly to one and by choosing $Z^2_0 = a_t A^*/a^2$  preserves the form of the mass terms,
\be
\frac{1}{N}\sum_x \sqrt{\widetilde g_x} = 1   \;  , \;   a_t  \sqrt{g_x} m^2 \phi^2_{tx} = \sqrt{\widetilde g_x} a^2 m^2 \widetilde  \phi^2_{tx}.
\ee
Introducing  a dimensionless mass ($m_0 = a m$) and coupling ($\lambda_0 = a^2 \lambda/Z^2_0$)  and dropping the {\em tilde} notation gives our lattice action in  Eq.~\ref{eq:FEMaction}. In our simulation, we also set the bare {\em speed of light} to one: $c_0 = a/a_t =1$. As a result, we have achieved the traditional advantage of setting all terms to  $\cO(1)$ in the lattice action, independent of the refinement.

Two consequences of our rescaling conventions should be noted.
First, the rescaled weights of the kinetic term give 
\be
\sum_{\< x,y \>} l^*_{xy}/l_{xy} = \frac{2}{3} (1  + \epsilon_0 ) E \; ,
\label{eq:kineticScale}
\ee
when summed over $E$ edges.  The $2/3$ factor is a consequence
of the equilateral triangulation  of the
icosahedron, each of which  contributes exactly 1 to the sum. However, when the triangles are projected obliquely onto  the  sphere they are no longer
equilateral so that the DEC weights increase the sum by a small geometrical fraction: $\epsilon_0 = 0.003285 + \cO(a^2/R^2)$. We also note that using  piecewise linear finite elements or the DEC, the  $\cO(a^2/R^2)$ corrections
in the classical action (\ref{eq:FEMaction}) are not determined. At present in the quantum context, we see no compelling advantage to higher order elements. 

Second, due to the intrinsic geometry of
the manifold, the action  still has explicit
lattice spacing dependence in the coefficient of the lattice Ricci scalar. In our rescaling convention, Eq.~\ref{eq:Rescaling}, this  coefficient is determined  relative to area of the sphere,
\be
a^2/R^2 = 4 \pi a^2/(A^* N ) =  8 \pi/(\sqrt{3} N)  \; .
\ee 
For the  non-perturbative CFT in the continuum
limit the Ricci term  is  an {\bf irrelevant operator} due
to the factor of $a^2/R^2$ combined with the known
scaling dimension $\Delta_\epsilon > 1$ for $\phi^2$. 
In our first extensive Monte Carlo simulations,
we have dropped the Ricci term in Eq.~\ref{eq:FEMaction} to demonstrate convergence to continuum CFT, as $a\rightarrow 0$. Although  the Ricci term is  not required, we 
 demonstrate subsequently the major advantage of including it is to accelerate  convergence to the continuum. 

\paragraph{\bf Restoring Symmetries in the QFT --} Now we  check  whether or not our classical FEM action in Eq.~\ref{eq:FEMaction},
is capable of converging to the full quantum $\phi^4$ theory   on $\mR \times \mS^2$. To this end, we search for a critical
surface in the bare coupling space $(m_0^2,\lambda_0)$ with extensive Monte Carlo simulations.  To search for the critical surface, we monitor the
4th order Binder cumulant,
\be
U_4(L,\mu_0, \lambda_0) =  \frac{3}{2} \left[ 1 - \frac{\< M^4\>}{ 3\< M^2\>^2} \right] ,
\ee
of  the magnetization $M = \sum_{t,x} \sqrt{g_x} \phi_{t,x}$,
as we reduce the lattice spacing $a$.
We impose periodic boundary conditions in $t$ with fixed aspect ratio, $L_t/L = 4$. 
In the continuum limit, our Binder cumulant is normalized so that it approaches $0$ in the extreme  disordered (Gaussian) phase and $1$ in the extreme ordered phase. A second order critical surface
is found where the Binder cumulant approaches
a constant between $0$  and  $1$ as the lattice spacing vanishes. We scan  the relevant parameter $m^2_0$ at fixed $\lambda_0 = 0.2$. 
\begin{figure}[h]
\centering
\includegraphics[width=\columnwidth]{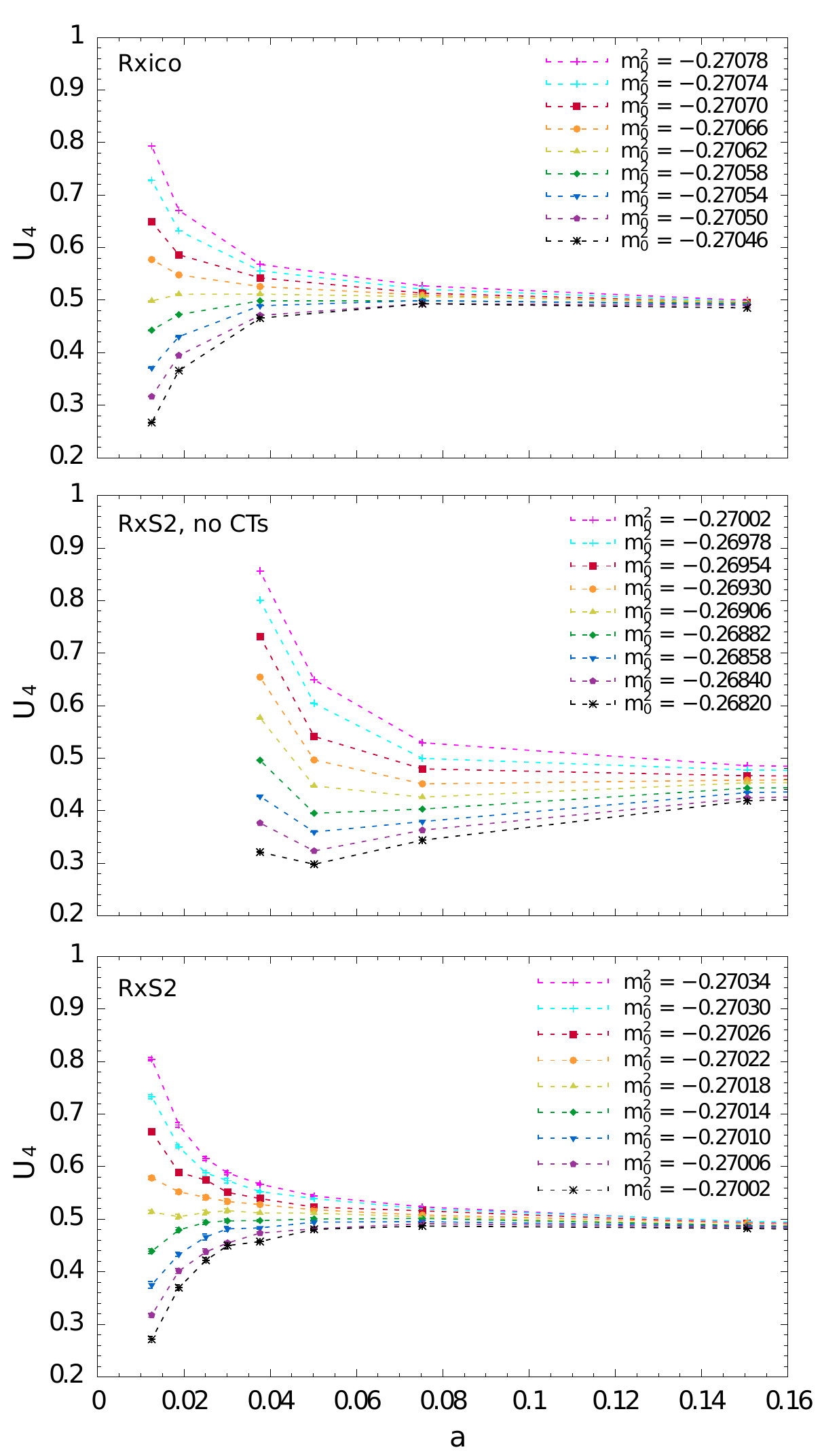} 
\caption{\label{fig:binder} Binder cumulant plotted against the lattice refinement for icosahedral spatial lattice (top)versus 
the spherical spatial lattice without  (middle) and with counter terms (bottom).  In all three cases, we searched in  $m^2_0$ at fixed $\lambda_0 = 0.2$}
\end{figure}

As a base line, in the top panel in Fig.~\ref{fig:binder}, we abandon the FEM weights by setting  $l^*_{xy}/l_{xy}=2/3$ and $\sqrt{g_x}=1$. As in our  radial 3d Ising simulation~\cite{Brower:2012vg} with icosahedron triangulations on $\mR \times \mIcos$,  there is apparently a well-defined critical theory. However  as shown below in Fig.~\ref{fig:symmetry}, the continuum fixed point exhibits only icosahedral irreducible multiplets, which break the full spherical symmetry at the level of  the $l=3$ descendant. The result is
presumably a CFT on an the icosahedron cylinder, $\mR \times \mIcos$, not our indented target manifold, $\mR \times \mS^2$.

In the middle panel in  Fig.~\ref{fig:binder},
we restore the  position dependent classical FEM weights  in Eq.~\ref{eq:FEMaction} for the simplicial lattice, but we now fail to locate a critical surface. For this study we have dropped the Ricci term and rescaled the fields to cancel $\epsilon_0$ in Eq.~\ref{eq:kineticScale}.
For values of $m_0^2 > -0.26906$, the cumulant trends towards $U_4=0$ for large $a$, but at smaller $a < 0.05$ the curves begin to turn around and will eventually oscillate. Ironically the FEM weights required for  classical physics  result in  failure for quantum physics. Just as in our 2d application~\cite{Brower:2018szu} on the Riemann Sphere, $\mS^2$, the problem is that the local FEM variations in the effective cutoff are amplified by the UV divergence
of the quantum field theory. To overcome this problem, we add  counter terms to our simplicial action, computed from spatially varying UV divergent lattice Feynman diagrams.  

\begin{figure}[h]
\includegraphics[width=\columnwidth]{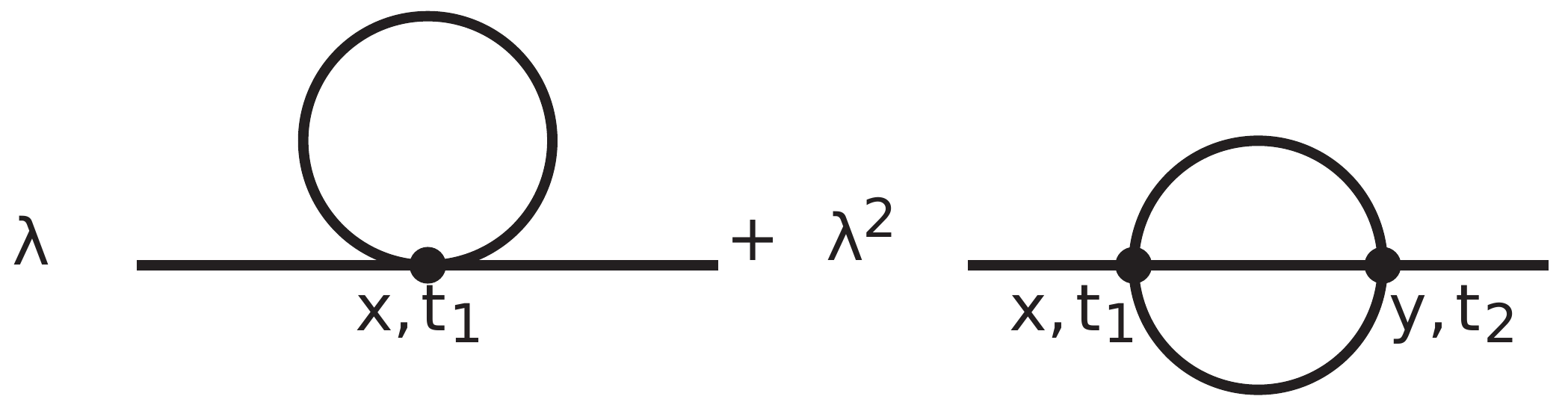}
 \caption{\label{fig:OneTwoLoop} The one- and two-loop UV divergent diagrams contributing to the mass renormalization in 3d. }
 \end{figure}

We need to define precisely how we convert the FEM action into the QFE action to compensate for  quantum UV defects. 
Since $\phi^4$  theory is super-renormalizable in three dimensions,  there  are a  finite number of divergent diagrams, illustrated in Fig.~\ref{fig:OneTwoLoop}: 
a one-loop linear divergence and a two-loop logarithmic divergence.  
To construct the simplicial Feynman diagrams in Fig.~\ref{fig:OneTwoLoop}, we compute 
numerically  the lattice propagator, $G_{t_1,x; t_2,y}$, for the  free theory at $m_0 = 0$ including
the Ricci term, which is required by conformal symmetry and to act as IR regulator on the sphere.
The result to second order is an effective action,  
\begin{flalign}
\Gamma_{eff} = \Gamma_0 + & 6 \lambda_0\sqrt{g_x}  G_{t,x; t,x}  \phi^2_{t,x} \\
 -& 24 \lambda^2_0 \sqrt{g_x} \phi_{t_1,x} G^3_{t_1,x; t_2,y}\ \phi_{t_2,y}  \sqrt{g_y}.  \nonumber
 \label{eq:effaction}
\end{flalign}
The counter terms will be designed to  exactly cancel
the rotational breaking in the relevant operators  in our lattice action.

Not  only is the  one-loop lattice diagram  local, it 
is also finite in lattice units.  This finiteness for
power divergences is a general feature of lattice perturbation theory. 
We subtract the rotational symmetric piece to isolate the breaking term,
\be
\delta G_{x}  \equiv G_{t,x;t,x} - \frac{1}{N} \sum_{x=1}^N \sqrt{g_x} G_{t,x;t,x} \; ,
\ee
which  is independent of $t$ by translation invariance a long the cylinder.
The two-loop  term  in Fig.~\ref{fig:OneTwoLoop} gives a non-local product $\phi_{t_1,x}\phi_{t_2,y}$. However, after subtracting the rotational symmetric piece, we find that the non-local behavior is exponentially damped in units of the lattice spacing -- or in  the jargon of lattice field theory, the correction is
local but not ultra-local. So to leading order in $a$, the dominant contribution  is  local and can be isolated by defining
\be
\delta G^{(3)}_x \equiv  \sum_{t',y} \sqrt{g_y} \big[ G^3_{t,x;t',y} - \frac{1}{N}
\sum_{x=1}^N \sqrt{g_x} G^3_{t,x;t',y} \big] \; .    
\ee
The sum over $t',y$ cancels a rotational symmetric logarithmic divergence, yielding a finite position dependent counter term as in the case of the one-loop counter term introduced for 2d $\phi^4$ theory on $\mS^2$. 
The reader is referred to Ref.~\cite{Brower:2018szu} for details. 

As a result, we propose a new  QFE action, 
\be
S_{QFE} = S - \sum_{t,x}\sqrt{g_x} [   6 \lambda_0  \delta G_{x} - 24 \lambda^2_0 \delta  G^{(3)}_x ] \phi^2_{t,x} \; ,
\label{eq:QFEaction}
\ee
relative to the classical FEM lattice
action, $S$, in Eq.~(\ref{eq:FEMaction}). By tuning to the weak coupling fixed point, this  QFE action  should  match the full renormalized perturbation following the Wilsonian renormalization procedure in the same fashion as on a regular hypercubic lattice.
Moreover we  see the new  QFE lattice action
in the bottom panel of Fig.~\ref{fig:binder}  does appear to have a  well defined critical surface.  Dropping the Ricci term in Eq.~\ref{eq:kineticScale}, the Binder cumulant is studied up to lattice sizes of $L=96$.  
We intersect the critical surface at $\lambda_0 = 0.2$ and $m^2_0 \simeq - 0.27018 (4)$, which we now fix as
a good approximation to the continuum critical couplings at zero
lattice spacing.

We now conjecture that our QFE lattice action (\ref{eq:QFEaction}), tuned to
the critical surface, also converges to the exact non-perturbative CFT as the cutoff is removed, which in the absence of a proof we support this  QFE conjecture with numerical simulations. While
this is the standard Wilson renormalization procedure used extensively on the hypercubic lattice in flat space, we acknowledge that this a non-trivial extension for our QFE action on curved  manifolds, which warrants further theoretical and higher precision numerical investigation. 
\begin{figure}[ht]
	\centering
	\includegraphics[width=\columnwidth]{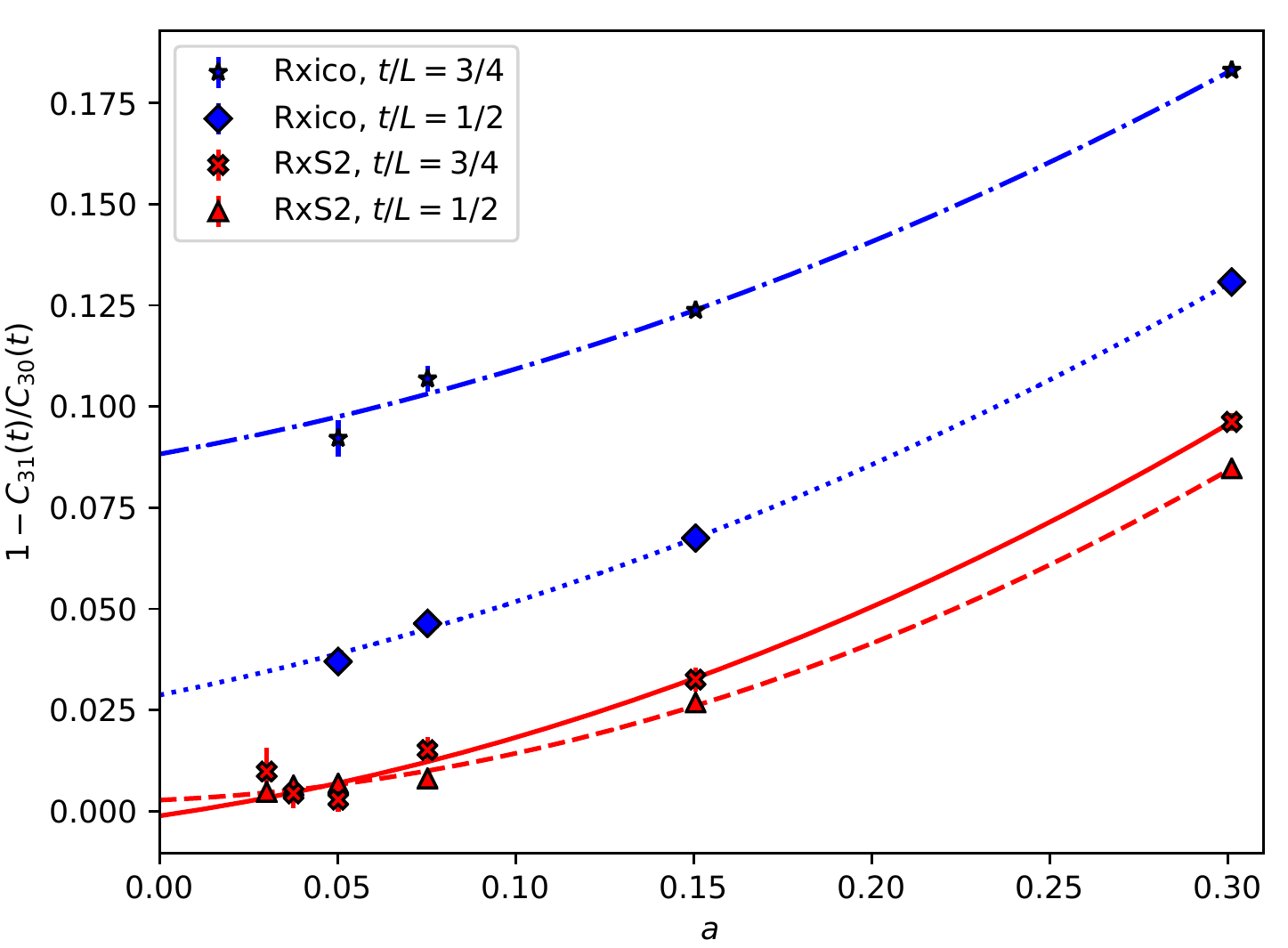}
	\caption{\label{fig:symmetry} 
Spherical symmetry breaking in the continuum limit for  the icosahedral model on $\mR \times \mIcos$ ({\bf Rxico}) compared with QFE on $\mR \times \mS^2$ ({\bf  RxS2}) for  the $l =3$ correlator at separations $t/L = 1/2, 3/4$. }
\end{figure}

Next, we check the restoration of $\text{SO}(3)$ spatial rotational symmetry by examining the  two-point correlator in the $\mathbb{Z}^2$-odd channel on the critical surface,
\be
C_{l,m_1,m_2}(t_1-t_2) = \< \phi_{t_1, l m_1} \phi_{t_2, l m_2} \>  \; , 
\label{eq:latticecorr}
\ee
projected onto partial wave on the sphere, $\phi_{t,lm} = \sum_{x} \sqrt{g_x}  \phi_{t,x} Y_{lm}[x]$.
The spherical harmonics are defined by evaluating the continuum  functions  at the discrete sites $x$.
In the continuum limit, spherical symmetry implies $C_{l,m_1,m_2}(t) =  \delta_{m_1, m_2} c_l(t)$,  with $2 l+1$ degeneracy
for $\text{SO}(3)$ irreducible representations. 
The first three $\text{SO}(3)$  representations, $\ell = 0,1,2$, are also irreducible in the icosahedral group. 
At the $l=3$ level, the seven dimensional $\text{SO}(3)$  representation
splits at finite lattice spacing into a direct sum of a three  (3T) and a four  (3G)  dimensional irreducible  representation 
in the icosahedral group.
We examine the diagonal elements $m_1=m_2=0$ and $m_1=m_2=1$ denoted by $C_{30}(t)$ and $C_{31}(t)$,  which are components entirely in the 3T and 3G icosahedral representations respectively.

In Fig.~\ref{fig:symmetry}, we  plot the normalized error in rotational symmetry, $1-C_{31}(t)/C_{30}(t)$, at two fixed physical time slices $t/L=1/2$ and $t/L=3/4$ as a function of the lattice spacing $a$.  A quadratic function is fit to each case to extrapolate to the continuum limit.  
Comparing the critical  theory on $\mR \times \mIcos$ and on $\mR \times \mS^2$, we see that without FEM weights and counter terms, the icosahedral breaking persists to the continuum limit as expected, whereas for our QFE action in Eq.~\ref{eq:QFEaction}, including the  one- and two-loop counter terms, the $\text{SO}(3)$ symmetry is recovered.

\paragraph{\bf Ricci improved Two Point Correlator -- } \label{sec:2pt}

Conformal symmetry completely determines the form of the two point function of primary operators,  which for a scalar in radial quantization is given by
\be
\< \phi(t_1, r_1) \phi(t_2, r_2)\> 
 =   \frac{1}{ (2 \cosh t_{12} - 2 \cos\theta_{12})^{\Delta_\phi}} \;,
\label{eq:2pt_function}
\ee
with $t_{12} = |t_1 - t_2|$ and $\cos\theta_{12} = r_1 \cdot r_2$  the relative coordinates on
the cylinder in Fig~\ref{fig:4pt}. 
Up to an overall normalization convention, the correlator is fixed by the scaling dimension $\Delta_\phi$ of the primary operator. 
Having  tested the restoration of spherical symmetry approaching the
critical surface at $ m^2_0 \simeq - 0.27018 (4)$ and $\lambda_0 = 0.2$, this correlator provides a stringent test of conformal symmetry for the continuum limit of the QFE lattice.

\begin{figure}[h]
	\centering
\includegraphics[width=\columnwidth]{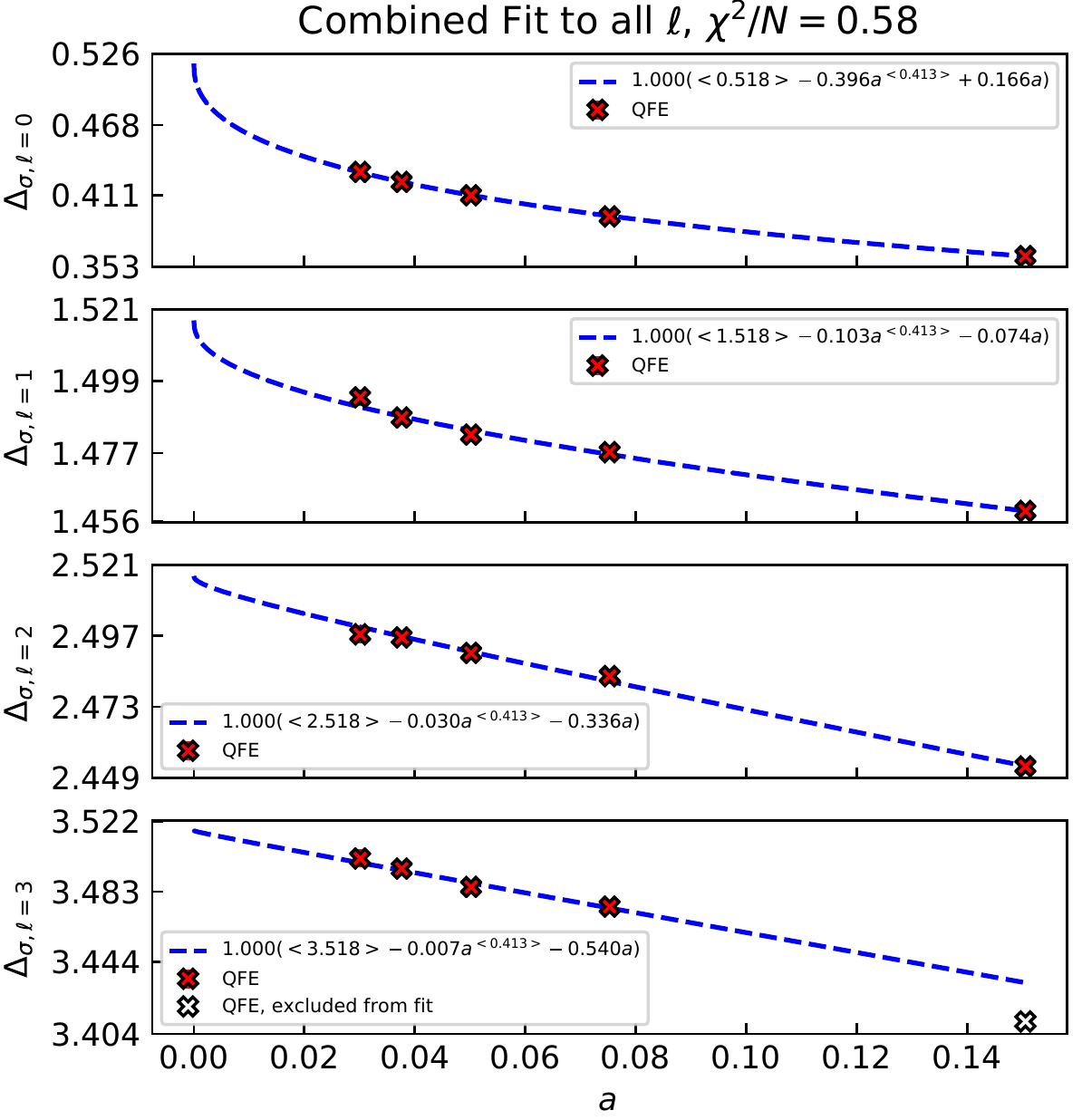}
	\caption{\label{fig:dims} Lowest $Z_2$ odd scaling dimension, $\Delta_{\sigma,\ell}$, as a function of lattice spacing, $a$, for the lowest four angular momentum values. 
	Quantities in angle brackets are held fixed in the continuum limit fits. Where not visible, statistical error bars are smaller than markers.} 
\end{figure}

The lowest primary in the $Z_2$-odd sector couples to the lattice field 
$\phi_{t,x}$ and must scale with the conformal dimension $\Delta_\sigma$ with unit spacing of its descendants $\Delta_{\sigma,\ell} =\Delta_\sigma + l$ for $l = 1,2,3, \cdots$ in the continuum limit.  
 The lattice correlator is projected onto partial waves on the sphere as  Eq.~\ref{eq:latticecorr} and fit to exponentials,  $\exp[ - \mu_l t_{12}]$, for large $t_{12} = |t_1 - t_2|$. 
The lattice exponential masses, $\mu_l$, are related to the scaling dimension by $\mu_l = a_t \; \Delta_{\sigma,l} = c_R \; a \; \Delta_{\sigma,l}$ in terms of a renormalized {\em speed of light} $c_R$ relative to the bare value $c_0$ which we set to 1 in the classical limit. We may fix the renormalized $c_R$ 
either by enforcing the integer spacing of descendants in the continuum limit or by matching the conserved dimension $\Delta_T = 3$ for the energy momentum tensor in the OPE expansion of 4-point function in Eq.~\ref{eq:expansion}.

The results  for $\Delta_{\sigma,\ell}$ as function of $a$ are given in Fig.~\ref{fig:dims}.  All the data is in a limited range of lattice spacing $a \in [0.03,0.15]$ corresponding to lattice sizes of $L\in [8,40]$ on periodic  cylinder with aspect ratio $L_t/L = 16$.
By computing the correlation function on $\cO(10^8)$ statistically independent configurations with an improved cluster  estimator~\cite{Brower:1989mt}, we achieve statistical errors for the scaling dimension on the order of 0.1\%. With such high statistics, our initial fits (not shown) clearly identified a lattice artifact term scaling as $\cO(a^{0.40})$ giving rise to an infinite slope as one approaches the continuum limit. This
is consistent at the 1\% level with the scaling dimension of the composite operator, $:\phi^2_{x,t}:$ ($\Delta_\epsilon = 1.4126$), implying   a cut off dependence, $\cO(a^{\Delta_\epsilon -1}) = \cO(a^{0.4126})$, when the Ricci term is not included
in the lattice action.   Indeed to
check this idea, we parameterize the fits shown in Fig.~\ref{fig:dims} to
\begin{equation}
\Delta_{\sigma,l}(a) = c \left(  \Delta_\sigma +l + A_l a^{ \Delta_\epsilon -1 } + B_l a \right) \label{eq:artifactfit}
\end{equation}
determining the constants $c$, $A_l$, and $B_l$  by a simultaneous fit to $l=0,1,2,3$ while setting
the dimensions to  their continuum values: $\Delta_\sigma\simeq 0.5181$ and $\Delta_\epsilon\simeq  1.4126$.
The linear term was  added to model subleading lattice artifacts beyond the irrelevant Ricci scaling.
For the coarsest lattice spacing at the $l=3$ level, we see that there may be sensitivity to yet further lattice artifacts, so we omit this point from the fit.

The result of the fit is shown by the blue dashed curves in Fig.~\ref{fig:dims}, with the best fit values of the coefficients reported in the legend.
We remark that the best fit value for the renormalized 
{\em speed of light} is very close to 1.
The small $\chi^2/N=0.58$ indicates that the Ricci term accurately captures the leading lattice artifact behavior, playing an important role in the continuum extrapolation despite being irrelevant. The correlator clearly converges to the conformal multiplet --- albeit slowly with the Ricci term excluded from the simulation --- as indicated by an accurate recovery of the descendent relation up to $l=3$ and consistency with the bootstrap value for $\Delta_\sigma$~\cite{Kos:2014bka}.

\begin{figure}[h]
	\centering
\includegraphics[width=\columnwidth]{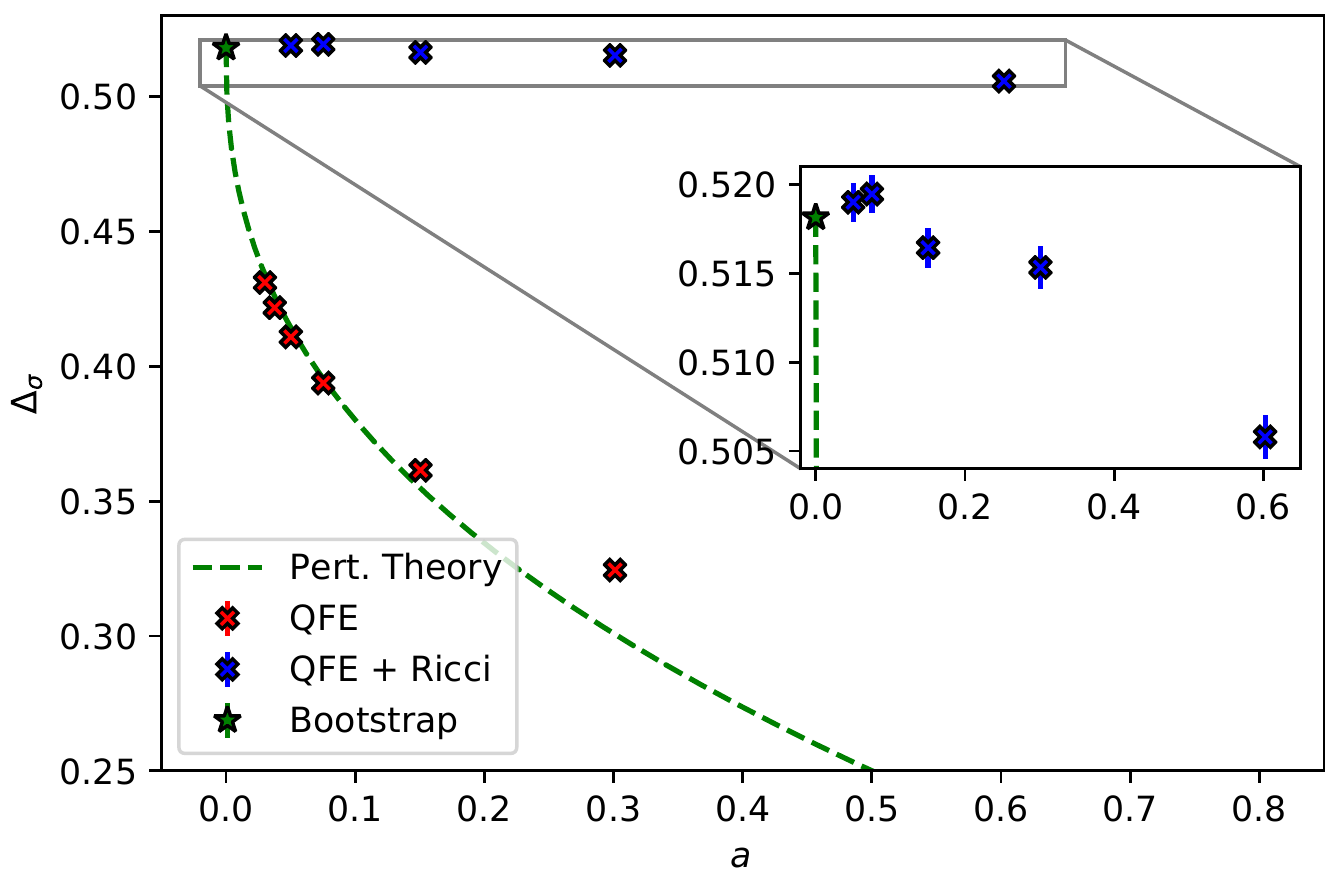}
	\caption{\label{fig:Rphi2} Scaling dimension of lowest $Z_2$ odd scalar primary $\sigma$ as a function of the lattice spacing, computed  with (blue crosses) and without (red crosses) the Ricci term in the QFE action.
	Fixing the $\Delta_\sigma = 0.5181$ at $a=0$ (green star) including  Ricci term in perturbation (green dashes)
	fits well the lattice spacing dependence. } 
\end{figure}

To check quantitatively the role of the Ricci term in approaching the continuum discovered in Fig.~\ref{fig:dims}, we carry out two additional studies.  First, we treat the Ricci term,  $\delta S = - (a^2/(8R^2))\sqrt{g_x}\phi_{t,x}^2$,  as a small perturbation near the continuum limit. To first
order this  shifts the scaling dimensions  by
\be
\delta \Delta_{\sigma,l} = - \frac{a^2}{8 R^2} \< \Delta_{\sigma,l} |\sqrt{g_x} \phi^2_{t,x} |\Delta_{\sigma,l}\>_c \; ,
\ee 
where the sum over $x$ is implied. The right hand side is independent of $t$ by translation invariance down the cylinder.
The form factor on the RHS remains a nonperturbative function of $\lambda_0$.  We have computed the matrix element  in our Monte Carlo simulation for $l=0$ and $l=1$ through a standard ratio method,
\be
\< \Delta_{\sigma,l} |\sqrt{g_x} \phi^2_{t,x} |\Delta_{\sigma,l}\>_c \simeq \frac{\langle \phi_{t_1,lm} \sqrt{g_x}\phi_{t,x}^2 \phi_{t_2,lm}\rangle_c}{ \langle \phi_{t_1,lm} \phi_{t_2,lm}\rangle} 
\ee 
for the connected piece, as $t_1/t_2 \rightarrow +/\hspace{-0.5em}- \infty$, which implements for the CFT the
standard {\em operator-state correspondence} map.
The resultant,
\begin{flalign}
\delta\Delta_{\sigma,l=0} &= -0.3777(38) a^{\Delta_\epsilon -1} \label{eq:pertartifact0} \\
\delta\Delta_{\sigma,l=1} &= -0.083(10) a^{\Delta_\epsilon -1}
\end{flalign}
are in  reasonable agreement with the coefficients found by fitting Eq.~\ref{eq:artifactfit} given in Fig.~\ref{fig:dims}. The perturbative  prediction in  Eq.~\ref{eq:pertartifact0}, shown in Fig.~\ref{fig:Rphi2} by the green dashed curve, has \emph{zero free parameters} after fixing to the  continuum value $\Delta_\sigma =0.5181$ at $a = 0$. The Ricci term describes the $a$ dependence of the data of the original simulation accurately at small $a < 0.2$.

Next we carried out a modest first simulation including the Ricci term in an improved QFE action for lattice spacings $a \in [0.05,0.6]$ on  $\cO(10^7)$ statistically independent configurations.
The blue markers in Fig.~\ref{fig:Rphi2} show the results from the simulation using the Ricci improved action compared to the red markers showing the original high statistics simulation omitting the Ricci term.
 The improved action simulation, highlighted in the insert, shows a dramatic improvement at lattice spacings $a < 0.2$, reducing finite lattice cutoff effects by two orders of magnitude and achieving an accuracy for the scaling dimension $\Delta_\sigma = 0.518(2)$ at about 0.5\% and in agreement with the bootstrap value.   In passing we note that for 2d $\phi^4$  on  $\mS^2$ \cite{Brower:2018szu}, this correction was not required, as expected since the  coefficient of the Ricci term for d = 2 vanishes.  

\paragraph{\bf Operator Product  Expansion --} \label{sec:ope}

The full content of a CFT requires computing data
for both the dimension of operators
and the 3-point coupling that  appears first in the 4-point function OPE.  Here we summarize our method to extract the
same parameters from QFE simulations on $\mR \times \mS^2$.
Radial quantization is well suited to study the OPE.

The invariant amplitude for identical scalars, $\phi_1 + \phi_2 \rightarrow \phi_3 + \phi_4$,  is
\be
g(u, v) = \frac{\< \phi_{t_1, r_1} \phi_{t_2, r_2}   \phi_{t_3, r_3}  \phi_{t_4, r_4}  \>}{ \<
  \phi_{t_1, r_1} \phi_{t_2, r_2} \> \< \phi_{t_3, r_3}  \phi_{t_4, r_4}  \>} \; ,
\ee
in terms of the standard cross ratios, $u$ and
$v$. In radial quantization it is more convenient to choose $\tau, \alpha$
coordinates,
\be
\cosh(\tau) = \frac{1 + \sqrt{v}}{\sqrt{u}} \quad,
\quad \cos(\alpha) = \frac{1 - \sqrt{v}}{\sqrt{u}} \; ,
\ee
where for large $\tau$ and fixed angle, $\sqrt{u}
  \simeq  4 \exp(- \tau)$.  To extract
  OPE terms, we place  two incoming fields on one sphere and two outgoing on a second separated by
$t = t_{12} - t_{34}$ as illustrated in  Fig~\ref{fig:4pt}.
Propagation is given by the dilatation operator $D$, so that the
invariant amplitude is 
\be
g(\tau, \alpha) =  \<|r_1 - r_2|^{2\Delta_\sigma} \phi_1 \phi_2 e^{ -t
  D} |r_3 - r_4|^{2\Delta_\sigma} \phi_3 \phi_4 \>, \nonumber
\ee
where the continuum two point functions have been inserted using the normalization convention in
Eq.~(\ref{eq:2pt_function}).   It should be noted that the ``time'' ($t$) separation along the cylinder is not in general a conformal invariant as it is conjugate to 
the dilatation operator. 

However, by choosing a special frame with antipodal points~\cite{Hogervorst:2013sma} on each unit sphere, the
time separation $t$ and angle $\theta$ are now mapped to
 invariants,
\be
\cosh(\tau)  = \cosh(t) \; , \; \cos(\alpha) =
  \hat r_1\cdot \hat r_3 =  \cos(\theta) \; 
\ee
and the   OPE expansion  for  $d=3$ is given as the partial waves  expansion,
\be
g(\tau, \alpha) 
= 1+\sum_{\Delta_l,
  l = 0,2,\cdots} \lambda^2_{\Delta_l} e^{ -\Delta_\ell t}  P_l(\cos\theta ) \; , 
  \label{eq:expansion}
\ee
where the leading contribution from the identity operator is normalized to be unity. Expanding 
conformal blocks into partial waves~\cite{Hogervorst:2016hal}, the 
couplings and  dimensions of the descendants are fixed by their primaries  and therefore restricting our
fitting parameters  to  CFT data.  (For $d\neq 3$, one replaces $ P_l(\cos\theta)$ by Gegenbauer polynomials, $C^{\nu}_\ell(\cos\theta)$, with $\nu=d/2-1$.)

 We are  particularly interested in the contribution from the stress-energy tensor, with $l=2$ and $\Delta_T=d$ to determine the central charge $C_T$.  Following Dolan and Osborn~\cite{Dolan:2000ut},  the coupling to the energy momentum tensor in Eq.\ref{eq:expansion} is 
\be
\lambda^2_{\Delta_T}  = \frac  {  4^{\Delta_T}\Delta_{\sigma}^2 \Delta_T^2}{ C_T d(d-1)} = \frac{96\Delta^2_{\sigma}}{C_T } \; .
\ee
Note in 2d 
the common practice is to use $c = C_T/2$ so the free field
is $c = 1$. We note in passing that the  bootstrap's xc-Minimization hypothesis  
for the central charge  gives  only a  few percent reduction
, $C_T/C_T^{free} =  0.946534(11) $, relative to the free theory
$C_T^{free} = d/(d-1) = 3/2$.  Indeed all the  dimensions of
the lowest primaries in the 3d Isings CFT and their couplings, with the exception of $\Delta_\epsilon$, are
within a few percent of the free (or generalized mean field) value,  so significant comparisons to
the bootstrap require high precision. 

To achieve this in  our future high precision simulations, the  Ricci QFE action will be used along with additional
improvement schemes common to numerical lattice field theory.  Since no lattice operator is a pure primary, we will  construct improved lattice sources to better approximate primary operators. As typical  example, in the lowest odd $Z_2$ scalar sector, with almost no additional cost an improved estimator in the cluster algorithm 
can measure a 2 by 2 matrix of 
correlator:  $C_{ij} = \<{\cal O}_i(t_1x_1) {\cal O}_j(t_1,x_2)\>$
for ${\cal O}_i(t,x)  = \{ \phi(t,x), \phi^3(t,x)\}$.
In principle,  as the lattice spacing
$a  \rightarrow 0$, the mixing is
$\cO((a/R)^{\Delta_{\sigma'} -  \Delta_\sigma})$. 
 By diagonalizing  the 2 by 2 matrix, the lowest state is an improved variational operator for $\sigma$ primary at finite lattice spacing. Using this operator will further reduce the lattice spacing errors in the  4pt function. 
 
\paragraph{\bf Future Directions --}\label{sec:conclusion}
Based on the numerical evidence in this letter, we believe
our  QFE lattice theory does converge to  
the continuum CFT as the cutoff is removed. The current simulation due to
the efficiency of the cluster algorithm generated $\cO(10^8)$ uncorrelated configurations
that allowed us to identify the importance
of including the Ricci term in the action.  Next
we plan to proceed to much higher precision numerical investigations to support the QFE lattice method against the best CFT data from the Conformal Bootstrap~\cite{El-Showk:2014dwa}.

Stringent comparison between QFE and the Conformal Bootstrap is interesting in its own right. Either agreement or disagreement could have fundamental consequences, since
the two  are based on radically different methods and assumptions. The QFE approach identifies a specific  target quantum field theory as the cutoff is removed, whereas the bootstrap tightly
constraints a generic class of theories by  combining 
exact inequalities on a truncated operator expansion 
with the observation that simple known theories nearly saturate the constraints, supplemented for example by the c-Minimization procedure~\cite{El-Showk:2014dwa}. 
Agreement between the two approaches would go a long way to support the validity of both. Disagreement will begin a
fundamental search for a better theoretical understanding  to hopefully remove the discrepancies.

By including fermions and gauge fields on the simplicial
complex, we believe QFE lattice theory should apply to
any  quantum field theory that posses  a renormalizable perturbative expansion on  a smooth Euclidean Riemann manifold~\cite{Luscher:1982wf}. Conformal
and special integrable  models provide stringent tests for QFE, but the general approach is equally applicable to  theories with massive deformations. The
mass deformation naturally induces a spectral flow from
eigenvalues of the Dilation operator  to the Hamiltonian, e.g.  operator dimensions ($\Delta$) to masses ($m$) respectively in the continuum limit: $1/R  \ll 
m  \ll 1/a$. QFE  is complimentary to the expanding repertoire of Hamiltonian Truncation~\cite{Rychkov:2014eea} and Lightcone Conformal Truncation~\cite{Anand:2020gnn} methods that also seek a non-perturbative computation
moving adiabatically away from  CFT fixed points.

Particularly interesting in this regard is the application of QFE to  4d non-Abelian gauge theories, which
are  under consideration as models for composite Higgs or dark matter~\cite{Brower:2019oor}, with enough fermionic flavors to be in or near the IR conformal window at strong coupling.  Supersymmetric conformal examples are also  under consideration. QFE methods should have an even wider range of applications to quantum gravity, for example in Anti-de-Sitter space~\cite{Brower:2019kyh} or the Regge formulation of simplicial gravity interacting with matter. None of these extensions are
easy or guaranteed to work, but we believe current success with 3d $\phi^4$ theory suggests a  way forward to test more complicated
field theories.

\paragraph{\bf Acknowledgements--}
We thank Casey Berger, Cameron Cogburn, Joel Giedt, Simeon Hellerman, Ami Katz, Kantaro Ohmori  and Domenico Orlando for useful discussions. This work was supported in part by the U.S. Department of Energy (DOE) under Award No. DE-SC0015845 for RCB and under Award No. DE-SC0019061 for GTF.  TR  acknowledges support from the Office of the Senior Vice President for  Research and Innovation at Michigan State University.

\bibliographystyle{unsrt}
\bibliography{3Dphi4}

\begin{thebibliography}{10}

\bibitem{Brower:2018szu}
Richard~C. Brower, Michael Cheng, Evan~S. Weinberg, George~T. Fleming,
  Andrew~D. Gasbarro, Timothy~G. Raben, and Chung-I Tan.
\newblock {Lattice $\phi^4$ field theory on Riemann manifolds: Numerical tests
  for the 2-d Ising CFT on $\mathbb{S}^2$}.
\newblock {\em Phys. Rev.}, D98(1):014502, 2018.

\bibitem{Brower:2016vsl}
Richard~C. Brower, Evan~S. Weinberg, George~T. Fleming, Andrew~D. Gasbarro,
  Timothy~G. Raben, and Chung-I Tan.
\newblock {Lattice Dirac Fermions on a Simplicial Riemannian Manifold}.
\newblock {\em Phys. Rev.}, D95(11):114510, 2017.

\bibitem{Brower:1989mt}
R.C. Brower and P.~Tamayo.
\newblock {Embedded Dynamics for $\phi^4$ Theory}.
\newblock {\em Phys.Rev.Lett.}, 62:1087--1090, 1989.

\bibitem{El-Showk:2014dwa}
Sheer El-Showk, Miguel~F. Paulos, David Poland, Slava Rychkov, David
  Simmons-Duffin, and Alessandro Vichi.
\newblock {Solving the 3d Ising Model with the Conformal Bootstrap II.
  c-Minimization and Precise Critical Exponents}.
\newblock {\em J. Stat. Phys.}, 157:869, 2014.

\bibitem{Cardy:1984rp}
John~L. Cardy.
\newblock {Conformal invariance and universality in finite-size scaling}.
\newblock {\em J. Phys.}, A17:L385--L387, 1984.

\bibitem{Cardy:1985lth}
J.~L. Cardy.
\newblock {Universal amplitudes in finite-size scaling: generalisation to
  arbitrary dimensionality}.
\newblock {\em J. Phys.}, A18(13):L757--L760, 1985.

\bibitem{Brower:2012vg}
R.~C. Brower, G.~T. Fleming, and H.~Neuberger.
\newblock {Lattice Radial Quantization: 3D Ising}.
\newblock {\em Phys. Lett.}, B721:299--305, 2013.

\bibitem{el2012solving}
Sheer El-Showk, Miguel~F Paulos, David Poland, Slava Rychkov, David
  Simmons-Duffin, and Alessandro Vichi.
\newblock Solving the 3d ising model with the conformal bootstrap.
\newblock {\em Physical Review D}, 86(2):025022, 2012.

\bibitem{Gasbarro:2019kgj}
Andrew~David Gasbarro.
\newblock {Studies of Conformal Behavior in Strongly Interacting Quantum Field
  Theories}.
\newblock Other thesis, 11 2019.

\bibitem{Arnold_2010}
Douglas~N. Arnold, Richard~S. Falk, and Ragnar Winther.
\newblock Finite element exterior calculus: from hodge theory to numerical
  stability.
\newblock {\em Bulletin of the American Mathematical Society}, 47(2):281–354,
  Jan 2010.

\bibitem{Kos:2014bka}
Filip Kos, David Poland, and David Simmons-Duffin.
\newblock {Bootstrapping Mixed Correlators in the 3D Ising Model}.
\newblock {\em JHEP}, 11:109, 2014.

\bibitem{Hogervorst:2013sma}
Matthijs Hogervorst and Slava Rychkov.
\newblock {Radial Coordinates for Conformal Blocks}.
\newblock {\em Phys. Rev.}, D87:106004, 2013.

\bibitem{Hogervorst:2016hal}
Matthijs Hogervorst.
\newblock {Dimensional Reduction for Conformal Blocks}.
\newblock {\em JHEP}, 09:017, 2016.

\bibitem{Dolan:2000ut}
F.~A. Dolan and H.~Osborn.
\newblock {Conformal four point functions and the operator product expansion}.
\newblock {\em Nucl. Phys.}, B599:459--496, 2001.

\bibitem{Luscher:1982wf}
M.~Luscher.
\newblock {Dimensional Regularization in the Presence of Large Background
  Fields}.
\newblock {\em Annals Phys.}, 142:359, 1982.

\bibitem{Rychkov:2014eea}
Slava Rychkov and Lorenzo~G. Vitale.
\newblock {Hamiltonian truncation study of the $\phi^4$ theory in two
  dimensions}.
\newblock {\em Phys. Rev.}, D91:085011, 2015.

\bibitem{Anand:2020gnn}
Nikhil Anand, A.~Liam Fitzpatrick, Emanuel Katz, Zuhair~U. Khandker, Matthew~T.
  Walters, and Yuan Xin.
\newblock {Introduction to Lightcone Conformal Truncation: QFT Dynamics from
  CFT Data}.
\newblock 2020.

\bibitem{Brower:2019oor}
Richard~C. Brower, Anna Hasenfratz, Ethan~T. Neil, Simon Catterall, George
  Fleming, Joel Giedt, Enrico Rinaldi, David Schaich, Evan Weinberg, and Oliver
  Witzel.
\newblock {Lattice Gauge Theory for Physics Beyond the Standard Model}.
\newblock {\em Eur. Phys. J.}, A55(11):198, 2019.

\bibitem{Brower:2019kyh}
Richard~C. Brower, Cameron~V. Cogburn, A.~Liam Fitzpatrick, Dean Howarth, and
  Chung-I Tan.
\newblock {Lattice Setup for Quantum Field Theory in AdS$_2$}.
\newblock 12 2019.

\end{thebibliography}
\end{document}